\documentstyle[sprocl,epsfig]{article}

\def\Journal#1#2#3#4{{#1} {\bf #2}, #3 (#4)}

\def\PRD{{\em Phys. Rev.} D}

\def\be{\begin{equation}}
\def\ee{\end{equation}}
\def\bea{\begin{eqnarray}}
\def\eea{\end{eqnarray}}

\begin{document}

\title{
\begin{bfseries}
\begin{boldmath}
STUDY OF $e^{+}e^{-} \rightarrow \tilde{\chi}^{+}_{1}
\tilde{\chi}^{-}_{1} \rightarrow \tilde{\tau}^{+}_{1}\nu_{\tau} +
\tilde{\tau}^{-}_{1}\overline{\nu}_{\tau}$
\end{boldmath}
\end{bfseries}
}

\author{YUKIHIRO KATO~\footnote{
This work is done in a collaboration with Mihoko M. Nojiri (Kyoto University), 
Keisuke Fujii (High Energy 
Accelerator Research Organization(KEK)),
and Teruki Kamon (Texas A\&M University). } }
\address{Department of Mathematics and Physics, Kinki University, \\
3-4-1 Kowakae, Higashi-Osaka, 577-8502, Japan}


\maketitle\abstracts{
In Supersymmetry with large $\tan\beta$ values,
the lighter stau ($\tilde{\tau}_{1}$) may be lighter than 
the lightest chargino ($\tilde{\chi}^{\pm}_{1}$).
The decay 
$\tilde{\chi}^{\pm}_{1} \rightarrow \tilde{\tau}^{\pm}_{1} \nu_{\tau}$,
followed by $\tilde{\tau}^{\pm}_{1} \rightarrow \tilde{\chi}^{0}_{1}\ \tau$,
alters a phenomenology of the chargino production.
We study the chargino pair production
at future $e^{+}e^{-}$ linear colliders.}
  
\section{Introduction}

Supersymmetry (SUSY) uniquely opens the possibility to directly
connect the Standard Model (SM) with an ultimate unification of the
fundamental interactions.
A primary goal for linear colliders is a precise measurement of
SUSY parameters if it would be discovered at Tevatron or LHC.
One of the most challenging signatures in SUSY will be events involving
$\tau$ leptons.
At large $\tan\beta$ values,
we could have a particular mass hierarchy of
$M(\tilde{\ell}) >  M(\tilde{\chi}^{\pm}_{1}) > M(\tilde{\tau}^{\pm}_{1}) > 
M(\tilde{\chi}^{0}_{1})$, because
a large mass splitting between $\tilde{\tau}_1$\
and other sleptons
($\tilde{e}$, $\tilde{\mu}$, $\tilde{\tau}_2$,  $\tilde{\nu}$) can be 
generated via $\tilde{\tau}_{L}-\tilde{\tau}_{R}$\ mixing.\cite{stau,stau2}\
Furthermore, 
Br($\tilde{\chi}^{\pm}_{1} \rightarrow \tilde{\tau}_{1} \nu_{\tau}$) = 100\%
if $\tilde{\chi}^{\pm}_{1}$ is gaugino-like.
Thus, the experimental signature is 2$\tau$'s
and missing energy and makes difficult to reconstruct the chargino mass.
We here study
$\tilde{\chi}^{+}_{1} \tilde{\chi}^{-}_{1} \rightarrow 
\tilde{\tau}^{+}_{1}\nu_{\tau} + \tilde{\tau}^{-}_{1}\overline{\nu}_{\tau}$\ 
in $e^+ e^-$ collisions within a framework of the Minimal Supersymmetric
Standard Model (MSSM)
by assuming
a reaction of 
$e^{+}e^{-} \rightarrow \tilde{\tau}_1^{+} \tilde{\tau}_1^{-}$,
followed by $\tilde{\tau}_1^{\pm} \rightarrow \tau^{\pm} \tilde{\chi}^{0}_{1}$,
is fully studied.\cite{stau,stau2}\

\section{Monte Carlo Simulation}

We assume 
(a)~$\tan\beta = 50$; 
(b) M$_{0} = 200$\ GeV, M$_{1}$ = 87.8 GeV and $\mu = -400$\ GeV; 
(c)~$\tilde{\tau}_{1}$\ is the next lightest SUSY particle (NLSP) 
        and decays to $\tilde{\chi}^{0}_{1} + \tau$;
(d)~$M(\tilde{\chi}^{\pm}_{1}) > 155\ {\rm GeV/c}^{2}$;
(e)~$M(\tilde{\chi}^{0}_{1})= 87\pm3\ {\rm GeV/c}^{2}$\ and 
      $M(\tilde{\tau}_{1}) = 152\pm4\ {\rm GeV/c}^{2}$,
      where two masses 
      are measured at center-of-mass energy 
      of $e^+ e^-$ collisions 
      ($e.g.$, $\sqrt{s}$ = 310 GeV)
      below the threshold of chargino pair production and
      the uncertainties~\cite{stau2} are estimated at 
      $\int {\cal L} dt = 100\ {\rm fb}^{-1}$;
(f) $\sqrt{s} = 400$\ GeV and $\int {\cal L} dt = 200\ {\rm fb}^{-1}$;
(g) electron beam is polarized at P($e^{-}$) = $-0.9$.
Four different SUSY points are studied:
$M(\tilde{\chi}^{\pm}_{1})$ = 157.5, 162.5, 172.5, and 192.5\ GeV/c$^{2}$.
The point with $M(\tilde{\chi}^{\pm}_{1})$ = 172.5\ GeV/c$^{2}$ refers to
minimal supergravity (mSUGRA) relations. 
The $\tau$\ polarization is 0.8 for all points.
With polarized  electron beam at P($e^{-}$) = $-0.9$,
the cross section for 
$e^{+}e^{-} \rightarrow \tilde{\chi}^{+}_{1}\tilde{\chi}^{-}_{1}$
is 20 times larger than that at P($e^{-}$) = $+0.9$.
Furthermore, the cross section for 
$e^{+}e^{-} \rightarrow \tilde{\tau}^{+}_{1}\tilde{\tau}^{-}_{1}$ is 
suppressed.

Two major backgrounds
for the 2$\tau$+missing energy final state are 
(i)~$e^{+}e^{-} \rightarrow
W^{+}W^{-} \rightarrow \tau^{+}\nu_{\tau} + \tau^{-}\overline{\nu}_{\tau}$
and 
(ii)~$e^{+}e^{-} \rightarrow \tilde{\tau}^{+}_{1}\tilde{\tau}^{-}_{1}
\rightarrow \tau^{+}\tilde{\chi}^{0}_{1} + \tau^{-}\tilde{\chi}^{0}_{1}$.
The two photon process and $Z^{0}Z^{0}$\ production have been studied
and found their contributions to be small.~\cite{stau}\
We consider only $WW$\ and $\tilde{\tau}^{+}_{1}\tilde{\tau}^{-}_{1}$\ 
production in the study.

As for the event generation, the helicity amplitudes for 
the process are calculated using the 
HELAS library.~\cite{HELAS}\
The final state $\tau$\ leptons are generated 
using the BASES/SPRING package~\cite{BASES}, 
and are decayed via TAUOLA
version 2.3.~\cite{TAUOLA} The effects of initial state radiation, 
beam energy spread, and beamstrahlung are also taken into 
account.~\cite{bespd}\ 

The cross sections for the signal events  
($M(\tilde{\chi}^{\pm}_{1}) = 172.5$\ GeV/c$^{2}$),
$\tilde{\tau}_{1}$\ pair production, and $W$\ pair production
($W^{-}W^{+} \rightarrow \tau^{-}\overline{\nu}_{\tau}+\tau^{+}\nu_{\tau}$) are
305~fb, 30~fb, and 256 fb at $\sqrt{s}=400$\ GeV.
We generate 100K events for each sample and pass them through
a JLC-type detector simulation.

\begin{figure}[htbp]
\vspace{-0.6cm}
\epsfysize=3.0in
\epsfxsize=3.0in
\center{\leavevmode\epsffile{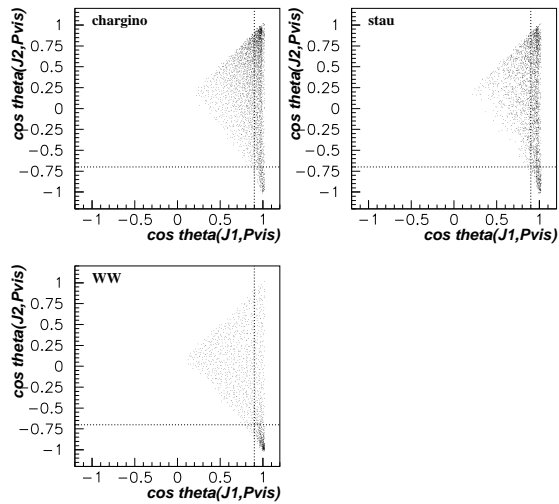}}
\vspace{-0.6cm}
\caption{
$\cos \theta(J_{2}, P_{vis})$ vs. $\cos \theta(J_{1}, P_{vis})$ 
for 
$\tilde{\chi}^{+}_{1}\tilde{\chi}^{-}_{1}$
(``signal''), $\tilde{\tau}^{+} \tilde{\tau}^{-}$ (``stau''), and
$W^+ W^-$ (``WW'') production.}
\label{fig:anglect}
\end{figure}

The event selection criteria are (1) two good jets with
$E\geq 5$\ GeV using JADE cluster algorithm (Y-cut $\geq 0.0025$), 
(2) jet angle cut: 
$\frac{-Q_{i}\cdot J_{zi}}{|\vec{J}_{i}|} \leq 0.8$
($Q_{i}$ and $\vec{J}_{i}$ are a charge and momentum of $i$-th jet,
where $i$ = 1 (2) for higher (lower) energy jet),
(3) $M_{jet} \leq 3$\ GeV/c$^{2}$,
(4) missing $P_{t} \geq 20$\ GeV/c, 
(5) $\cos \theta(J_{1}, P_{vis}) \leq 0.9$\ or 
$\cos \theta(J_{2}, P_{vis}) \geq -0.7$,
(6) Thrust $\leq$ 0.98,
and (7) Acoplanarity angle $\geq 30^{\circ}$.
In cut (5), $\cos\theta(J_{i}, P_{vis})$\ is defined to be 
$\frac{\vec{J}_{i}\cdot \vec{P}_{vis}}
{\vert\vec{J}_{i}\vert\vert\vec{P}_{vis}\vert}$.
Here $\vec{P}_{vis}$ is the visible momentum vector.
This cut can separate $WW$\  background from signal events. 
Figure~\ref{fig:anglect} shows the distributions of this variable for each
process. 
The $WW$\ event concentrates a region, $\cos \theta(J_{1}, P_{vis}) 
\geq 0.9$\ and $\cos \theta(J_{2}, P_{vis}) \leq -0.7$. 
Since 
the $\tau$ in $W \rightarrow \tau \nu$
remains going to the $W$ boson direction,
we can effectively reject the $WW$ background using this requirement.

\section{Result}

Table~\ref{tab:eventsel} shows the results of the efficiencies
after each cut for signal event at mSUGRA point and two background processes. 
\begin{table}[htbp]
\caption{Result of selection efficiencies}
\label{tab:eventsel}
\begin{center}
\begin{footnotesize}
\begin{tabular}{|l|r|r|r|}
\hline 
process & $\tilde{\chi}^{+}_{1} \tilde{\chi}^{-}_{1}$ & $\tilde{\tau}^{+} 
 \tilde{\tau}^{-}$ & $W^{+}W^{-}$ \\
(P($e^{-}$)= $-0.9$) &(ref) & &
($\rightarrow \tau^{-}\nu_{\tau}\tau^{+}\nu_{\tau}$) \\
$\sigma$ (fb) & 305 & 30 & 256 \\
\hline
\hline
(1)\ 2 good jets & 0.849 & 0.866 & 0.793 \\
(2)\ Jet angle cut & 0.581 & 0.625 & 0.233 \\
(3)\ $M_{jet} \leq 3$ GeV/c$^{2}$ & 0.538 & 0.596 & 0.227 \\
(4)\ Missing $P_{t} \geq 20$ GeV/c & 0.427 & 0.467 & 0.168 \\
(5)\ $\cos\theta (J_{1},Pvis) \leq 0.9$ or & & & \\
\ \ \ \ \ $\cos\theta (J_{2},Pvis) \geq -0.7$ & 0.403 & 0.426 & 0.090 \\
(6)\ Thrust $\leq 0.98$ & 0.399 & 0.420 & 0.075 \\
(7)\ Acoplanarity $\geq 30^{\circ}$ & 0.365 & 0.369 & 0.053 \\
\hline
\hline
$\epsilon$ &  0.365 & 0.369 & 0.053 \\
\hline
\# of expected events & & & \\
for $\displaystyle{\int {\cal L} dt}$ = 200 fb$^{-1}$ 
& 22265 & 2214  & 2714  \\
\hline
\multicolumn{4}{r}{($M(\tilde{\chi}^{\pm}_{1})=172.5\ {\rm GeV/c^{2}},
M(\tilde{\tau}^{\pm}_{1})=152.5\ {\rm GeV/c^{2}}$)} \\
\end{tabular}
\end{footnotesize}
\end{center}
\end{table}
Since $\tau$\ from $W$\ decay escapes to the forward region, 
$W^{+}W^{-}$\ events are significantly rejected by the jet angle cut.
Cuts (5) and (6) are also effective to suppress $W^{+}W^{-}$\ events.

\begin{figure}[htbp]
\vspace{-0.6cm}
\epsfysize=2.5in
\epsfxsize=2.9in
\center{\leavevmode\epsffile{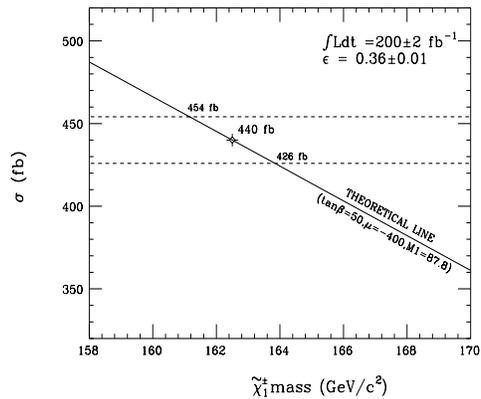}}
\vspace{-0.6cm}
\caption{$\sigma(e^{+}e^{-} \rightarrow
\tilde{\chi}^{+}_{1}\tilde{\chi}^{-}_{1} \rightarrow
\tilde{\tau}^{+}_{1}\nu_{\tau} + \tilde{\tau}^{-}_{1}
\overline{\nu}_{\tau}$) vs. $\tilde{\chi}^{\pm}_{1}$\ mass}
\label{fig:final3}
\end{figure}

The event acceptances for the various chargino mass points are shown in 
Table~\ref{tab:eventsel1}.
\begin{table}[htbp]
\caption{Event acceptance for four different chargino masses where
$M(\tilde{\tau}_{1})$ = 152.5 GeV/c$^{2}$\ and $M(\tilde{\chi}^{0}_{1})$
= 86.9 GeV/c$^{2}$.}
\label{tab:eventsel1}
\begin{center}
\begin{footnotesize}
\begin{tabular}{l|rrrr}
\hline
MC Pt. & 1. & 2. & 3. & 4. \\
       &    &    & (ref) & \\
\hline
$M(\tilde{\chi}^{\pm}_{1})$ (GeV/c$^{2}$) & 157.5 & 162.5 & 172.5 & 192.5 \\
$\Delta M$ (GeV/c$^{2}$) & & & & \\
$(\equiv M(\tilde{\chi}^{\pm}_{1}) - M(\tilde{\tau}))$ & 
5.0 & 10.0 & 20.0 & 40.0 \\
\hline
$\sigma$ (fb) & 486 & 440 & 305 & 111 \\
$\epsilon$ & 0.36 & 0.36 & 0.37 & 0.38 \\ 
\hline
\multicolumn{5}{r}{($\mu = -400$, $\tan\beta = 50$, $\tau$\ polarization 
= 0.8)} \\
\multicolumn{5}{r}{($P(e^{-}) = -0.9$)} \\ 
\end{tabular}
\end{footnotesize}
\end{center}
\end{table}
We find that the event acceptance is independent
of the  $\tilde{\chi}^{\pm}_{1}$\ mass, 
because the mass difference between $\tilde{\tau}_{1}$\ and 
$\tilde{\chi}^{0}_{1}$\ determines a $\tau$-jet energy 
and $\tilde{\chi}^{0}_{1}$\ energy determines the missing energy in a event.
Therefore it might be possible to determine the cross section without 
$\tilde{\chi}^{\pm}_{1}$\ mass 
information. We estimate the accuracy of the production cross-section 
measurement using the event acceptance and the integrated luminosity. 
Assuming the accuracy
of the luminosity measurement is 1\% ($200\pm2$\ fb$^{-1}$) and 
the event acceptance measurement is 3\% ($0.36\pm0.01$), we calculate the
accuracy of the production cross-section. This calculation takes into account
the statistical uncertainty.  
Figure~\ref{fig:final3} shows the 
result of this calculation. 
The accuracy of the production cross-section measurement can be 
3.2\%.

\begin{figure}[htbp]
\vspace{-0.6cm}
\epsfysize=2.7in
\epsfxsize=2.7in
\center{\leavevmode\epsffile{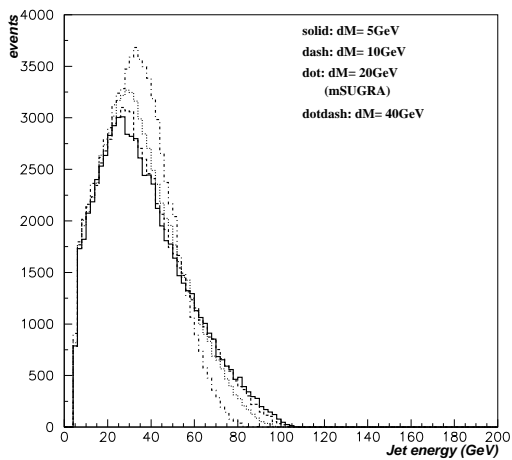}}
\vspace{-0.6cm}
\caption{$\tau$-jet energy distributions for various $\Delta M$\ values
(without normalization by luminosity)}
\label{fig:final2}
\end{figure}

\begin{figure}[htbp]
\vspace{-0.6cm}
\epsfysize=2.7in
\epsfxsize=2.7in
\center{\leavevmode\epsffile{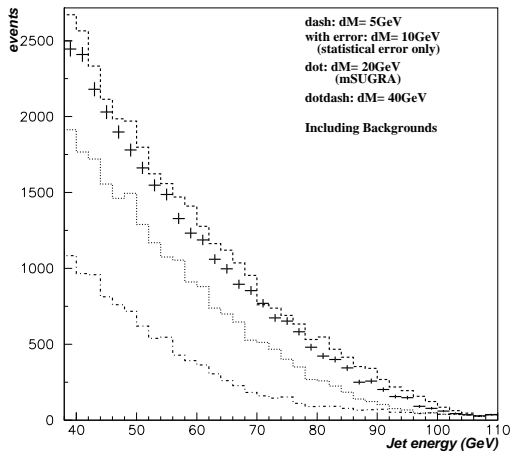}}
\vspace{-0.6cm}
\caption{$\tau$-jet energy distributions around end points for 
$\int {\cal L} dt = 200\ {\rm fb}^{-1}$}
\label{fig:final}
\end{figure}

An endpoint of the $\tau$-jet energy distribution strongly depends on the mass
difference, $\Delta M \equiv M(\tilde{\chi}^{\pm}_{1}) - 
M(\tilde{\tau}_{1})$. 
Figure~\ref{fig:final2} shows the $\tau$-jet energy distributions
for the various $\Delta M$\ values before luminosity normalization. 
We see the end point decreases as $\Delta M$\ increases.
There is a possibility
of the $\tilde{\chi}^{\pm}_{1}$\ mass determination using this distribution
without any other information. Figure~\ref{fig:final} shows the
$\tau$-jet energy distribution (including background events) at
$\int {\cal L} dt = 200\ {\rm fb}^{-1}$. 
We also show the result on $\Delta M = 10\ {\rm GeV/c}^{2}$\ with a
statistical uncertainty expected at 200~fb$^{-1}$. We should be able to
measure the $\tilde{\chi}^{\pm}_{1}$\ mass better than 5 GeV/c$^{2}$ 
using the $\tau$-jet distribution. 

\section{Conclusion}

We studied $\tilde{\chi}^{+}_{1}\tilde{\chi}^{-}_{1}
\rightarrow \tilde{\tau}^{+}_{1}\nu_{\tau} + \tilde{\tau}^{-}_{1}
\overline{\nu}_{\tau}$ in  $e^{+}e^{-}$ collisions,
where the stau is the NLSP.
The process is quite possible in particular at large $\tan\beta$\ values 
and completely
alters a phenomenology of lighter chargino pair production.
$W^{+}W^{-}$\ production, which will be a major SM background, can be
suppressed by kinematical cuts on (a) angle between $\vec{P}_{vis}$\ and
$\vec{J}_{i}$ and (b) thrust of the event.
With $\int {\cal L} dt = 200\ {\rm fb}^{-1}$\
at $\sqrt{s} = 400$ GeV, 22000 signal events
for our reference point ($M(\tilde{\chi}^{\pm}_{1})=172.5\ $GeV/c$^{2}$) 
and 5000 background events pass our selection cuts.
Using these samples, we can measure
the cross section for chargino pair production better than 3.2\% and
the chargino mass better than 5 GeV/c$^{2}$.

\section*{Acknowledgments}

I acknowledge the organizing committee for successful run
of this workshop.

\end{document}